\documentclass[12pt,preprint]{aastex}
\usepackage{natbib}

\shorttitle{Heat Conduction in Cooling Flows}
\shortauthors{Soker, Blanton, \& Sarazin}
\begin{document}

\title{Cooling of X-ray Emitting Gas by Heat Conduction
in the Center of Cooling Flow Clusters}

\author{Noam Soker\altaffilmark{1,2},
Elizabeth, L. Blanton\altaffilmark{2,3},
and
Craig L. Sarazin\altaffilmark{2}}

\altaffiltext{1}{ Department of Physics, Technion $-$ Israel Institute of Technology,
Haifa 32000, Israel;
soker@physics.technion.ac.il.}

\altaffiltext{2}{Department of Astronomy, University Craig L. Sarazin,
P. O. Box 3818, Charlottesville, VA 22903-0818;
eblanton@virginia.edu, sarazin@virginia.edu.}

\altaffiltext{3}{Chandra Fellow.}

\begin{abstract}

We study the possibility that a large fraction of the
gas at temperatures of $\sim 10^7$ K in cooling flow clusters
cools by heat conduction to lower temperatures, rather than by
radiative cooling.
We argue that this process, when incorporated into the so-called
``moderate cooling flow model'', where the effective age of the
intracluster medium is much lower than the age of the cluster,
reduces substantially the expected X-ray luminosity from gas
residing at temperatures of $\lesssim 10^7$ K.
In this model, the radiative mass cooling rate of gas at
$\sim 10^7$ K inferred from X-ray observations, which
is $< 20 \%$ of the mass cooling rates cited in the past,
is easily met.
The heat conduction is regulated by reconnection between
the magnetic field lines in cold ($\sim 10^4$ K) clouds and the
field lines in the intracluster medium.
A narrow conduction front is formed, which, despite the relatively
low temperature, allows efficient heat conduction from the hot ICM
to the cold clouds.
The reconnection between the field lines in cold clouds and those
in the intracluster medium occurs only when the magnetic field
in the ICM is strong enough.
This occurs only in the very inner regions of cooling flow clusters,
at $r \sim 10-30$ kpc.
The large ratio of the number of $H\alpha$ photons to the
number of cooling hydrogen atoms is explained by this scenario.
\end{abstract}


\section{Introduction} \label{sec:intro}

Recent {\it Chandra} and {\it XMM-Newton} observations appear to be
in contradiction with the simplest cluster cooling flow models.
While hot, $T \simeq 1-8 \times 10^7$ K, X-ray emitting gas
seems to cool at rates of $10-500 \, M_\odot$ yr$^{-1}$, there
is little indication for gas cooling below a temperature
of $\sim 10^7$ K (e.g., Molendi \& Pizzolato 2001;
Kaastra et al.\ 2001; Peterson et al.\ 2001,
2002, 2003; Fabian et al.\ 2002a).
Namely, the limit on the cooling rate below $\sim 10^7$ K
inferred from X-ray observations is $< 20 \%$ of the
mass cooling rates cited in the past (Fabian 2002),
but compatible with the moderate cooling flow model of
Soker et al.\ (2001).
On the other hand, X-ray observations reveal properties that
seem to support the cooling of gas.
First, there is no indication for violent ongoing heating
(e.g., shocks or cluster  mergers) to offset radiative cooling.
Hot bubbles of radio plasma may supply heat, but they also
compress the gas, which reduces its cooling time
(Soker et al.\ 2001).
High spatial resolution X-ray images from {\it Chandra} reveal complex
X-ray structures in the inner $\sim 30$ kpc of cooling flow clusters,
(e.g., McNamara 2002 and references therein),
including X-ray deficient bubbles that rise buoyantly transporting energy
from the AGN to the intracluster medium (ICM).

Morris \& Fabian (2003) suggest that the weakness of the X-ray emission
lines below $\sim 10^7$ K can be at least partially reconciled with larger
cooling rates
if most of the metals are concentrated in a small fraction ($\sim 10$\%)
of the cooling gas.
Soker et al.\ (2001) suggest that the cooling flow model can be
put on a more solid ground if some assumptions are relaxed,
e.g., the cooling flow age is taken to be much below the age
of the clusters (Allen et al.\ 2001; Bayer-Kim et al.\ 2002).
These relaxed assumptions lead to much lower mass cooling rates,
in what was called the ``moderate cooling flow model''.
In the present paper, we continue along these lines, and show that
moderate cooling flow models can be compatible with the non-detection
of line emission from cooler gas if some of the cooling hot X-ray
gas in moderate cooling flow models is cooling from
$\sim 10^7$ K to $10^4$ K by heat conduction.
Heat conduction has similar
effects to those of mixing, as proposed by, e.g.,
Fabian et al.\ (2001, 2002a) and Bayer-Kim et al.\ (2002).
Observations show that magnetic fields become strong at the
center of cooling flow clusters (Eilek \& Owen 2002, and references
therein).
Magnetic fields inhibit heat conduction across field lines.
We propose that reconnection of magnetic field lines occurs between
hot X-ray emitting gas and cooler gas at $\sim 10^4$ K.
We argue that the conduction fronts are narrow, making heat conduction
very efficient.

 We note here that for strong magnetic fields,
the self-inhibiting of the heat-flux along magnetic field lines
due to scattering of conducting electrons off plasma waves
is not efficient
(Pistinner \& Eichler 1998).
When the magnetic field is weak, heat conduction is
suppressed along field lines (Levinson \& Eichler 1992), while
when the magnetic field is strong, the suppression is small
(Pistinner \& Eichler 1998).
Pistinner, Levinson, \& Eichler (1996) consider this effect in
cooling flow clusters.
They show that in the outer parts where
the thermal pressure is much higher than the magnetic pressure,
the heat conduction is suppressed by this mechanism by
one to three orders of magnitude, while in the inner parts it is
not suppressed much if the magnetic field is strong (see their Figs. 1-3;
eq. 69 of Pistinner \& Eichler 1998 shows that modest suppression may occur).
In addition to that effect, immediately after reconnection, heat
conduction may be quite efficient in the turbulent region
(Lazarian \& Cho 2003).
We therefore assume here that heat conduction is efficient along
magnetic field lines in regions where the magnetic field is
close to equipartition.

Our magnetic reconnection model is different from
papers dealing with large scale ($\gtrsim 100$ kpc) magnetic fields
and conduction between exterior very hot intracluster gas and
hot gas within the cooling flow region
(e.g., Bregman \& David 1988; Pistinner \& Shaviv 1996;
Norman \& Meiksin 1996; Narayan \& Medvedev 2001; Loeb 2002;
Voigt et al.\ 2002;
Fabian, Voigt, \& Morris 2002b;
Ruszkowski \& Begelman 2002; Markevitch, Vikhlinin, \& Forman 2002).
In many of these models, heat conduction heats the gas in the cooling
flow region, and may reduce the cooling rate or even eliminate it completely.
Other studies argued that heat conduction is
suppressed by a large factor, and hence plays no role in the cooling
flow process.
We instead examine local heat conduction at lower temperatures, after the
gas has lost most of its thermal energy.
The gas continues to cool to lower temperatures, but does so
conductively, rather than radiatively.
Since it cools rapidly due to conduction, we do not see strong low
ionization X-ray lines
(e.g., Peterson et al.\ 2001)
or evidence for lower temperature X-ray multiphase gas
(e.g., Molendi \& Pizzolato 2001)
in the X-ray spectra of cooling flow regions in clusters.

\section{The Proposed Scenario}
\label{scenario}

The basic scenario is as follows.
Gas cools radiatively from the ambient cluster temperature to $\sim 10^7$ K,
and slowly streams inward, becoming gravitationally bound to the cD galaxy.
As a result of the cD galaxy motion through the ICM
and radio bubble activity, there is a constant mixing of gas in the
inner regions ($r \lesssim 10-30$ kpc) of cooling flow clusters.
Different parcels of gas collide with one another.
In particular, we will assume that cold ($T \la 10^4$ K) clouds move
rapidly through the hotter ICM.
Some processes involving cold clouds in cooling flows, among them
heat conduction, were studied before (Loewenstein \& Fabian 1990;
Ferland, Fabian, \& Johnstone 2002).
We note that colder material is known to exist
in many cooling flow clusters,  as inferred from optical filaments
(see \S~\ref{sec:optical} below),
and from recent CO observations conducted by
Edge (2001; see also Edge \& Frayer 2003, and Salome \& Combes 2003)
who found the molecular mass to be in the range of
$\sim 10^9-10^{11.5} M_\odot$.
These cold clouds may result in large part from stripping of the ISM
from galaxies and subclusters.
However, our scenario is in accord with the presence of a moderate rate
of cooling of intracluster gas, and this may produce all or part of
the cold gas.
Evidence for cooling flows still exists in the {\it Chandra} and
{\it XMM/Newton} observations, but with much lower cooling rates
than previously suggested
(e.g., Bayer-Kim et al.\ 2002; David et al.\ 2001; Blanton et al.\ 2003).
The cooling gas may cool to a temperature of $\sim 10^4$ K
and below.
In particular, the heat conduction process itself cools the hot gas,
forming more cold clouds, although it can also evaporate the cold
clouds.
The sign of the conductive mass flux (evaporation or cooling?) is
discussed further in \S~\ref{reconnect}.

The observed line widths of the optical emission line clouds in cooling
flows are typically a few hundred km s$^{-1}$
(e.g., Heckman et al.\ 1989).
Similar line widths are seen for the molecular hydrogen clouds seen in C0
(e.g., Edge 2001).
Since we identify the cold clouds in our model with these observed
materials (see \S~\ref{sec:optical} below),
we will assume that the cold clouds move through the ambient hot gas at
speeds of a few hundred km s$^{-1}$.
These motions are mildly subsonic in the ambient gas, and are comparable or
slightly greater than the velocity dispersions of stars in the same central
regions of the cD galaxies.
Thus, it is likely that these motions are orbital or infall velocities,
perhaps supplemented by other dynamical effects (e.g., the expansion of
radio bubbles).

We note that in the moderate cooling flow model, the mass
cooling rate evolves with time, being zero at the the beginning
(Soker et al.\ 2001; their discussion following eq. [4.17]).
Therefore, for young clusters the number of clouds from cooling gas
is small (cold clouds originating from striped galaxies may still
be present), but so is the mass cooling rate. Therefore, even in
young clusters the X-ray emission from gas at $T \lesssim 10^7$~K
should be weak on average.

As a result of cooling and inflow, the density of the
ICM increases by a factor of $\sim 10$.
An isotropically-tangled magnetic field in the gas will have its pressure
increased by a factor of $\sim 10^{4/3}$.
Radial inward motion (Soker \& Sarazin 1990), stretching of
ICM magnetic fields by expanding jets and
bubbles (Soker 1997), as well as a possible cluster
dynamo (Godon, Soker, \& White 1998),
may result in a much larger increase in the magnetic pressure.
A dynamically important magnetic field in the inner regions
of cluster cooling flows is inferred from observations
(Eilek \& Owen 2002; Taylor, Fabian, \& Allen 2002).
The enhanced magnetic fields and the gas motions in the inner
cooling flow region will most likely result in reconnection of magnetic
field lines.
The energy released by reconnection can play an important role in the
energetics of the optical filaments
(Sabra, Shields, \& Filippenko 2000), via heating produced directly by
reconnection (Jafelice \& Friaca 1996; Godon, Soker \& White 1998)
and/or by Alfv\'en waves (Friaca et al.\ 1997), and/or by facilitating heat
conduction (B\"ohringer \& Fabian 1989).

At present, we are mainly interested in reconnection
between the magnetic fields in cold clouds ($\sim 10^4$ K) and in
the hot ($\sim 10^7$ K) ICM.
Such reconnection can enhance heat conduction from the hot gas into
the cold clouds, and this may result in the hot gas cooling conductively
rather than radiatively.
That is, the thermal energy released in cooling the hot gas is delivered
to the cold clouds, and is radiated as optical/UV line emission at
$\sim 10^4$ K.
Replacing radiative cooling of the hot gas with conductive cooling
will reduce the emission at low but X-ray emitting temperatures,
and could account for the weak soft X-ray luminosity from gas
at intermediate temperatures $10^6 \lesssim T \lesssim 10^7$ K.
In addition to allowing conductive heat flow, reconnection may also
cause mixing of the gas via the fast flow emanating from the reconnection
site (Lazarian \& Cho 2003).
Mixing was also suggested as a mechanism to suppress X-ray
emission lines below $\sim 1$ keV (e.g., Oegerle et al.\ 2001;
Fabian et al.\ 2001, 2002a; Johnstone et al.\ 2002).
Fabian et al.\ (2002a) briefly mention that heat conduction
to cold gas plays a similar role to that of mixing.
Recently, Cho et al.\ (2003) have argued that magnetic fields are
relatively
ineffective at suppressing turbulent heat conductions.
It is unclear how turbulent in the regions around the clouds will be;
the existence of large-scale abundance gradients in cooling flows indicates
that global turbulence is not very efficient
(B\"ohringer et al.\ 2004).

In principle, reconnection might break the field
lines connecting hot and cold gas, rather than reconnect them,
and this would suppress thermal conduction.
Here, we assume that the field lines in the cold clouds are initially
disconnected from the field lines in the hot ICM, as much as the magnetic field
lines of the Earth are disconnected from the magnetic field lines
of the Solar wind.
We assume that reconnection only occurs
near the stagnation point at the front of the cloud
where ram pressure is maximum, and
in the tail behind the motion of the cloud.
A schematic description of the evolution of a single
ICM magnetic field line is given in Figure 1.
This is similar to what is observed in the Earth-Solar wind interaction
(e.g., Oieroset et al.\ 2001).
For this configuration, heat conduction substantially increases
after the reconnection event.

One worry with this scenario might be that reconnection near the
stagnation point at the front of the cloud would be too slow to occur
before the magnetic fields in the ambient hot gas were advected downstream
(in the frame of the cloud).
Exactly at the stagnation point, the relative velocity is zero, but
the relative velocity returns to its full value as the ambient gas
moves around the cloud.
Thus, magnetic fields lines might slip around the cloud before they
reconnect.
In strongly magnetized plasmas with magnetic field reversals, reconnection
can occur at a fraction of the Alfven speed
(e.g., Petschek and Thorne 1967).
As noted above, the speed of the cloud relative to the ambient gas is a
fraction of the sound speed in the ambient gas.
At the front of the cloud near the stagnation point, this motion will
be somewhat slower.
Thus, there is (just) enough time for magnetic fields to reconnect at
the front of the cloud.
It is also possible that viscosity or Kelvin-Helmholtz instabilities will
help to keep magnetic fields in the ambient gas close to the clouds for
a somewhat longer period.

As drawn, Figure 1 shows a smooth laminar reconnection of the
magnetic field lines behind the cloud.
If there is a turbulent wake, the geometry will be
more complex, but the same basic process may occur.
What is crucial is that the field lines from the clouds are
eventually disconnected from the ICM.
A turbulent wake may even enhance the rate of reconnection.
In any case, the stretched field lines will overcome
the turbulence pressure and reconnect.

Rayleigh-Taylor and Kelvin-Helmholtz instability modes are
expected to occur in the front interface between the cloud and the
ICM.
It is likely that the Kelvin-Helmholtz instabilities will be partly
suppressed by the strong magnetic fields
(e.g., Vikhlinin, Markevitch, \& Murray 2001).
In any case, the breakup of clouds by instabilities may not be a
problem for the present scenario, since we require small clouds.
In the next section the size range which is optimal for the proposed
scenario is derived.

One concern with this argument is raised by the {\it Chandra}
observations of ``cold fronts'' in clusters
(see the review by Markevitch et al.\ 2002).
A cold front is a contact discontinuity observed in many
clusters of galaxies between a large region of relatively cool gas
($\sim 10^7$ K) which is moving rapidly through the hot ICM.
In some cases, the cold fronts are the cooling cores
associated with a merging subcluster.
The sharp temperature jump observed in cold fronts implies that the
magnetic field there practically inhibits heat conduction.
Since the geometry is essentially identical to the geometry of cold clouds
moving through hot ICM, why doesn't magnetic field line reconnection
occur in cold fronts and lead to rapid thermal conduction?
The answer may lie in the relative strength of the magnetic field at
the stagnation point at the front of the cold front or cold cloud.
In the proposed scenario, the magnetic field in the cold clouds
is more or less in pressure equipartition with the thermal gas,
and it is quite strong, reaching $\sim 10 \%$ of equipartition,
in the ICM.
The magnetic field in either sides of cold fronts may
be increased by stretching and reach a value of $\sim 10 \mu{\rm G}$,
as inferred from the requirement to suppress instabilities
(Markevitch et al.\ 2002), but it is still below equipartition values.
 Note that we are examining reconnection in the ambient
post-shock and compressed gas which has a slow speed relative to
the cloud.
In this case it is the magnetic pressure that drives the reconnection,
and not a direct fast collision of magnetic field lines.
In this case the condition for fast reconnection to occur
is that the magnetic pressure be comparable or larger than
the thermal pressure (Petschek \& Thorne 1967; see discussion
in section IIb of Soker \& Sarazin 1990).

\section{Magnetic Field Reconnection and Heat Conduction}
\label{reconnect}

In this section we quantify some of the arguments, and estimate the
importance of some of the processes, discussed in the previous section.
We start by considering the rate of heat conduction across the
interfaces between hot ICM and cold cloud gas in the inner
regions of cooling flows.
Our treatment of the evolution of these conduction fronts follows
Borkowski, Balbus, \& Fristrom (1990), which should be consulted
for a more detailed treatment.
Heat conduction proceeds on a time scale of
$\tau_{\rm con} \simeq ( 5 n {\rm k} T/2)(\chi T^{7/2} L^{-2})^{-1}$,
where the first parenthesis is the enthalpy per unit volume, and the
second parenthesis is the heat loss per unit volume per unit time
because of heat conduction.
Here, $\chi$ is (nearly) a constant defined such that $\chi T^{5/2}$
is the heat conduction coefficient, $n$ is the total number density,
and $L$ the width of the conduction front between the cold cloud
and the ICM in its surrounding.
Substituting typical values, we obtain
\begin{equation}
\tau_{\rm con} \simeq 4 \times 10^5
\left( \frac {n_e} {0.04 \, {\rm cm}^{-3}} \right)
\left( \frac {T} {10^7 \, {\rm K}} \right)^{-5/2}
\left( \frac {L} {0.1 \, {\rm kpc}} \right)^{2} \, {\rm yr} \, ,
\label{eq:tcon}
\end{equation}
where $n_e$ is the electron density.
Basically, $\tau_{\rm con}$ is the time required to cool the
hot phase gas located inside the conduction front via heat conduction.
Our scenario requires that this time be much
shorter than the radiative cooling time (see eq. \ref{eq:volu} below).
For the conduction time to be short, the conduction front must remain
narrow.
However, the heat conduction front thickness does increase with time.
After a time $t_{\rm br}$, the conduction front thickness is (Balbus 1986)
\begin{equation}
L(t_{\rm br}) \simeq 0.15
\left( \frac{t_{\rm br}}{10^6 \, {\rm yr}} \right)^{1/2}
\left( \frac{n_e}{0.04 \, {\rm cm}^{-3}} \right)^{-1/2}
\left( \frac{T}{10^7 \, {\rm K}} \right)^{5/4} \, {\rm kpc} \, .
\label{eq:Levol}
\end{equation}
The isobaric radiative cooling time in the temperature range
$2 \times 10^5 \lesssim T \lesssim 4 \times 10^7$ K is
(see cooling curve in, e.g., Gaetz, Edgar, \& Chevalier 1988)
\begin{equation}
\tau_{\rm rad} \simeq 2 \times 10^8
\left( \frac {n_e} {0.04 \, {\rm cm}^{-3}} \right)^{-1}
\left( \frac {T} {10^7 \, {\rm K}} \right)^{3/2} \, {\rm yr} \, .
\label{eq:trad}
\end{equation}

To explain the low luminosity of the X-ray emission lines for gas in the
temperature range $10^6 \lesssim T \lesssim 5 \times 10^6$ K by heat
conduction, we require $\tau_{\rm con} < \tau_{\rm rad}$ at these
temperatures.
This translates into a condition on the width of the conduction front.
Let us first consider the non-magnetic case.
Assuming the gas is isobaric
and using equations~(\ref{eq:tcon}) and (\ref{eq:trad}),
we find the condition to be
\begin{equation}
L < 0.28 \left( \frac { T_{\rm min} } { 5 \times 10^6} \right)^{3}
\, {\rm kpc} \, .
\label{eq:Lcon}
\end{equation}
Here,
$T_{\rm min}$ is the temperature below which the gas
must cool conductively to avoid detection by X-ray line emission.
Thus, heat conduction dominates at those lower temperatures as long
as heat conduction occurs in very narrow fronts, with a scale much
shorter than the size of the cooling flow region.

Equation~(\ref{eq:Lcon}) is a necessary condition for conduction to
dominate for radiative cooling, but it is not sufficient.
This condition guarantees that conduction dominates over radiative cooling
within the narrow conduction fronts.
However, although conductive cooling only occurs within these narrow
fronts, radiative cooling goes on everywhere within the cooling flow
region.
Thus, while conduction might dominate locally, radiative cooling could
be more important overall, and the resulting soft X-ray lines might
still exceed the observational limits.
Let $V_{\rm con}$ be the total volume of the hot gas within conduction
fronts (the total volume cooling via heat conduction) at any given time.
The conductive cooling time of the hot gas in this volume is
$\tau_{\rm con}$
(eq.~\ref{eq:tcon}).
Let $V_{\rm CF}$ be the total volume of the whole cooling flow region,
where the isobaric radiative cooling time is $\tau_{\rm rad}$
(eq.~\ref{eq:trad}).
We assume that $V_{\rm con} \ll V_{\rm CF}$.
The condition that conduction cooling dominate radiative cooling
for the gas at temperatures of $\lesssim T_{\rm min}$ is that
on average at all times
\begin{equation}
\frac {V_{\rm con}}{\tau_{\rm con}}>
\frac {V_{\rm CF}}{\tau_{\rm rad}}.
\label{eq:volu}
\end{equation}
For the last equation to hold the following should be true in
the environment of cooling flow clusters.
First, the conduction at these low temperatures must be fast,
which requires narrow conduction fronts (eq.~\ref{eq:Lcon}).
 We note here that since we assume the presence of many clouds, the
width of the conduction front around any cold cloud can be much smaller
than the overall size of the inner cooling flow region, $L \ll 20$ kpc.
Second, the clouds should be small, so that the cooling fronts are narrow,
and there should be a large number of them so that a large enough volume
is covered, according to equation~(\ref{eq:volu}).
Third, in order to cool the hot gas via conduction, the magnetic field
lines of the cold clouds should be reconnected to those of the ICM.
This requires, as stated earlier, reconnection at the head of the cloud.

Usually it is expected that heat conduction is more important at
high temperatures than at low temperatures.
Thus, one might not expect conduction to play an important role in the
coolest X-ray emitting gas detected in clusters, as we are suggesting.
It will, however, occur for cool gas if the conduction fronts are narrow.
Also, if the heat conduction is regulated by magnetic reconnection,
then it may occur preferentially in cooler gas.
This might occur because the magnetic pressure can become dynamically
important in the cooler inner region.
If the magnetic field is isotropically turbulent, then the magnetic pressure
grows with density as $P_B \propto n_e^{4/3}$.
The ratio of the magnetic to thermal pressure then varies as
$P_B/P_{\rm th} \propto n_e^{1/3} T^{-1}$.
The lowest temperature inferred from X-ray emission in CF clusters
is typically $T_{\rm min} \sim T_c/3$, where $T_c$ is the cluster temperature
outside the cooling flow region.
The density in the inner region where the temperature is $T_{\rm min}$
is $\sim 10$ times higher than in the region where the temperature is $T_c$.
Since in an inflow the magnetic pressure can increase much faster than
as $n_e^{4/3}$ (Soker \& Sarazin 1990), the ratio $P_B/P_{\rm th}$
is probably increased by a factor of $\ga 10$ in going from the
general ICM to the inner cooling flow region.
Observations also support the presence of strong magnetic
fields in the centers of cooling flow clusters (Eilek \& Owen 2002).
The stretching of ICM magnetic field lines around the cold clouds
may increase the ratio $P_B/P_{\rm th}$ by an additional factor.
This may be sufficient to allow reconnection between the magnetic
field in the ICM and that inside the fast moving clouds.
We argue that the requirement for reconnection sets the
lower temperature in CF clusters.
The magnetic pressure in the ICM is still slightly below the equipartition
value, and ram pressure and stretching  due to the motions of cold clouds
may amplify the field to achieve equipartition near the fronts of the
clouds.
In this scenario, the negligible heat conduction
across cold fronts is because the magnetic field is below
its equipartition value, as discussed earlier.

 From the discussion above (eq.~\ref{eq:Lcon}),
we conclude that the typical cloud sizes should be $L \lesssim 0.1$ kpc.
We consider such small cold clouds moving through the ICM.
Reconnection is likely to occur in two places, as is seen in the
interaction of the Earth's magnetic field with the Solar wind
(e.g., Oieroset et al.\ 2001).
At the leading edge of the cloud, the magnetopause, the magnetic
field of the ICM and the cloud's magnetic field are compressed
together, leading to reconnection
if the field lines are not aligned.
As the cloud moves relative to the ICM, the field lines are stretched
behind the cloud, and eventually reconnect at a distance of several cloud
radii behind the cloud.
For example, in numerical magnetohydrodynamic simulations of the
interaction of a cloud with a moving medium, Gregori et al.\ (2000)
find that the reconnection region extends to 5-10 cloud radii.
Although the results of Gregori et al.\ (2000) do not correspond
one-to-one with the conditions in clusters
(e.g., the clouds are not magnetized in their simulations),
we note that they show that magnetic effects tend to be important
only if the initial Mach Alfvenic number is sufficiently small.
This applies to the situation studied here.
Once reconnection occurs and leads to conductive cooling, part of the
ICM may be accreted onto the cold cloud or may form another cold cloud.
In either case, the frozen-in magnetic fields will be accreted as
well, and this will lead to more field lines connecting the hot ICM
with cold clouds, and to more conduction.
Moreover, the compression of the ICM fields during cooling will lead
to very strong magnetic fields within the cold clouds.

In this geometry, and based on the numerical simulations
of Gregori et al.\ (2000), the heat conduction from the ICM to
the cloud occurs along the field lines which enclose an ICM
volume only $\Gamma \sim 10$ times larger than the cloud volume at
each moment.
For the proposed mechanism to work, we must, however, postulate
a much more efficient reconnection, with an ICM volume of
$\Gamma \sim 100$ times larger than the cold cloud volume.
With a temperature ratio of $\sim 10^3$ and pressure equilibrium,
this means that the
ICM mass is $\sim 10^{-3} \Gamma$ times the cloud mass.
The radiative coefficient $\Lambda$ at $10^7$ K is about 10 times that
for ionized gas at $10^4$ K,
so the ratio of the luminosity of the cold cloud to that of the ICM
connected to it is $L_{\rm cold}/L_{\rm hot} \sim 10^{5}/ \Gamma \sim 10^3$.
Most of the luminosity comes in the optical and UV, rather than
in the X-ray band.
Without magnetic field lines enclosing the interaction region,
clouds with a size smaller than the Field length will evaporate
(e.g., McKee \& Begelman 1990).
In the present magnetic field geometry, on the other hand,
the ICM is being cooled rather than the cloud being evaporated.
This is because at any given time the total mass of the gas in the
hot phase which is connected to the cloud is $\sim 0.1$ times
the mass of the cold cloud.
 (Recall our estimate above, based on the simulations of
Gregori et al.\ 2000, that only a small volume of the surrounding
gas is magnetically connected to the cold cloud).
Thus, even without any radiative
cooling, thermal equilibrium will bring the total mass enclosed within
the magnetic field lines to a temperature of only $\lesssim 0.05$ times the
temperature of the hot phase.
Thus, there is not enough thermal energy in the hot phase in contact with
the clouds to evaporate them.
Again, some non-negligible fraction of the hot-phase energy
is still expected to be radiated in the X-ray band. From
equation (\ref{eq:tcon}), the cold cloud will cool the region
connected to it in $\sim 5 \times 10^5$ yr.
In a time $t$ years, the clouds will cool a mass of
$\sim 10^{-3} (t/5\times 10^5) \Gamma M_{\rm cold}$,
where $M_{\rm cold}$ is the total mass of the cold clouds
in the inner region.
The cooling rate via heat conduction is then
\begin{equation}
\dot M({\rm conduction}) \sim 20
\left( \frac{\Gamma}{100} \right)
\left( \frac{M_{\rm cloud}}{10^8 \, M_\odot} \right)
\, M_\odot \, {\rm yr}^{-1} \, ,
\label{eq:dotmh}
\end{equation}
with strong dependence on the width and geometry of the conduction front.
For example, parcels of hot-phase gas cooling first, will
cool via heat conduction parcels of gas still hot and on
the same field lines, i.e., the freshly cooling gas will serve
as cold clouds.
If large masses of cold dense gas are present in cooling flow clusters
(Ferland et al.\ 2002), then a significant amount of gas can cool via
heat conduction.
As evident from the last equation, however, we still expect the
conductive cooling rates to be much smaller than the radiative cooling
rates suggested by pre-{\it Chandra} and
{\it XMM-Newton} X-ray observations,
and to be consistent with the amounts of cold matter and rates of
star formation observed in the centers of cooling flows
(e.g., Soker et al.\ 2001).
The heat conduction cooling rate may become a significant, and even
dominant, fraction of the cooling rate of the hot phase in the
moderate cooling flow model (Soker et al.\ 2001).

The implication is that the cooling ICM
mainly loses its thermal energy below $\sim 1$ keV via heat conduction
rather than line emission.
Some line emission is still expected.
Mixing can also lead to a reduction in the radiative cooling rate
(Bayer-Kim et al.\ 2002), and
was discussed in a turbulent environment by Begelman \& Fabian (1990).

To summarize, combining the moderate cooling flow model and
the heat conduction process proposed here,
the total line emission expected for gas cooling from $\sim 10^7$ K
to $\sim 10^6$ K might be factor of $\sim 30-100$
lower than that expected in older versions of the cooling flow model,
where only line emission is considered for the cooling process, and
where the cooling rates are much higher.
This factor comes from a $\sim 10$ times lower mass cooling rate of
the hot gas (Soker et al.\ 2001), and from another factor of $\sim 3-10$
for cooling via heat conduction.
This much lower expected radiative cooling rate at low
X-ray emitting temperatures is compatible with limits
inferred from observations.
This limit is $< 20 \%$ of the expected cooling rate in
old versions of the cooling flow model (Fabian 2002).
Of course, energy conservation implies that eventually the gas cools
by radiation.
This occurs at much lower temperatures, $\sim 10^4$ K.
Therefore, the large ratio of the number of $H\alpha$ photons to the
number of cooling hydrogen atoms is explained by this scenario,
as well as by mixing (Bayer-Kim et al.\ 2002, their section 8).
 From energy conservation, the total luminosity of optical/UV/IR emission
from the cold gas is
\begin{eqnarray}
L_{\rm opt} & \approx & \frac{5}{2} \, \frac{k T_{\rm min}}{\mu m_p} \,
\dot{M_X} \nonumber \\
& \approx & 7 \times 10^{42}
\left( \frac{T_{\rm min}}{10^7 \, {\rm K}} \right)
\left( \frac{\dot{M_X}}{30 \, M_\odot \, {\rm yr}^{-1}} \right)
\, {\rm erg} \, {\rm s}^{-1} \, .
\label{eq:lopt}
\end{eqnarray}
The fraction of this luminosity which comes out in any one line depends
on the abundances, temperatures, densities, and excitation conditions
in the gas, but roughly one expects a few percent of the luminosity
to emerge in H$\alpha$ (e.g., Heckman et al.\ 1989).
Thus, the expected H$\alpha$ luminosities due to conductive heating are
\begin{equation}
L({\rm H}\alpha) \sim 2 \times 10^{41} \left( \frac{T_{\rm min}}{10^7 \,
{\rm K}} \right)
\left( \frac{\dot{M_X}}{30 \, M_\odot \, {\rm yr}^{-1}} \right)
\, {\rm erg} \, {\rm s}^{-1} \, .
\label{eq:lHalpha}
\end{equation}
This is crudely comparable to the optical line luminosities observed from
cooling flows (e.g., Heckman et al.\ 1989).

\section{Correlation between Optical and X-ray Emitting Gas in Cooling Flow
Clusters}
\label{sec:optical}

A strong spatial correlation between bright regions of X-ray emitting gas
and optical ($H\alpha$) filaments has been found with {\it Chandra}
observations of several cooling
flow clusters including Hydra A (McNamara et al.\ 2000), Abell 1795 (Fabian
et al.\ 2001), Abell 2052 (Blanton et al.\ 2001), and Virgo/M87
(Young et al.\ 2002).
One of the clearest correlations of bright X-ray and $H\alpha$ emission is in
Abell 2052, where the $H\alpha$ emission corresponds with the brightest parts
of the shells surrounding bubbles evacuated by the cluster's central radio
source.  A temperature map of Abell 2052 (Blanton et al.\ 2003) shows that
these regions have cooled to about 0.8 keV ($9\times10^{6}$ K), or about 1/4
of the cluster temperature outside of the cooling flow region.  The
$H\alpha$ emission represents gas at a temperature of approximately $10^4$ K,
and it is possible that cooling between $\approx 10^7$ and $10^4$ K
has occurred by conduction.

Currently, the best measurements of mass-deposition rates in cooling flow
clusters can be made with high-resolution grating spectroscopy using
{\it Chandra} and {\it XMM-Newton}.  These observations are able to resolve
individual emission lines expected from gas emitting at a range of temperatures.
Peterson et al.\ (2003) present results from 14 cooling flow clusters
observed with the Reflection Grating Spectrometer on {\it XMM-Newton}.  The
mass-deposition rates are measured from $kT$ to $kT/2$, $kT/2$ to $kT/4$,
$kT/4$ to $kT/8$, and $kT/8$ to $kT/16$, where $kT$ is the temperature of
the cluster outside of the cooling flow region.  Mass-deposition rates
($\dot{M}$)'s were
successfully measured for 12 of the 14 clusters, and of these 12, 9 of them
had well-constrained $\dot{M}$'s in the $kT$ to $kT/2$ and $kT/2$ to
$kT/4$ ranges.  For the remaining three, only upper limits were possible in
the $kT/2$ to $kT/4$ range.
We have used the $\dot{M}$'s in the $kT/2$ to $kT/4$ range from Table 5
in Peterson et al.\ (2003), and taken $kT/4$ as the limiting temperature
for each cluster.  For all clusters that have measured $H\alpha$ luminosities
(Heckman et al.\ 1989, Allen 1995),
we have calculated the expected $H\alpha$ luminosity ($L_{H\alpha}$) from
cooling by conduction using equation \ref{eq:lHalpha}.  We were able to make
this comparison using ten of the clusters.  A comparison of the predicted
vs.\ the measured $H\alpha$ luminosities is presented in Table 1, and plotted
in Figure 2.  The error bars represent 1-$\sigma$ errors calculated by the
propagation of errors (of the temperature and mass-deposition values).
The dotted line represents $L_{H\alpha}$(measured) $= L_{H\alpha}$(predicted).
Points fall on both sides of the line, in other words, equation
\ref{eq:lHalpha} sometimes over-predicts and sometimes under-predicts the
amount of $H\alpha$ emission.  Still, the points generally follow the line
with higher predicted $L_{H\alpha}$ values corresponding with higher
measured $L_{H\alpha}$ values.  This is consistent with the correlation
of $\dot{M}$ and measured $L_{H\alpha}$ described in Heckman et al.\ (1989).

\section{Summary} \label{sec:conclusion}

We proposed that the upper limit on the mass cooling rate of gas
cooling below a temperature of $\sim 10^7$ K in cooling flow clusters
(e.g., Kaastra et al.\ 2001; Peterson et al.\ 2001, 2002, 2003)
can be made compatible with the existence of cooling flows in
the following way.
First, one must consider the so called ``moderate cooling flow model''
(Soker et al.\ 2001), where the effective age of the ICM is much
lower than the age of the cluster.
This implies a much lower mass cooling rate, by a factor of $\sim 10$,
than that in the old version of the cooling flow model.
Second, the possibility that a large fraction of the
gas at $\sim 10^7$ K cools by heat conduction to lower
temperatures, rather than by radiative cooling, further reduces the
X-ray luminosity expected from gas below $T\sim 10^7$ K.
With these two processes, the limit on the cooling rate below
$\sim 10^7$ K inferred from X-ray observations, which
is $< 20 \%$ of the mass cooling rates cited in the past
(Fabian 2002), is easily met.

In the present paper we showed that reconnection between
the magnetic field lines in cold ($\sim 10^4$ K) clouds and the
field lines in the ICM, if it occurs, forms a narrow conduction front.
Despite the relatively low temperature, the narrow conduction front
allows efficient heat conduction from the hot ICM to the cold clouds.
We also showed that (eq. \ref{eq:dotmh}) a large fraction of the ICM
at $\sim 10^7$ K will cool by heat conduction.
The reconnection between the field lines in cold clouds and those in the
ICM occurs only when the magnetic field in the ICM is
strong enough, i.e., close to the equipartition value.
This occurs only in the very inner regions of cooling flow cluster,
at $r \sim 10-30$ kpc.

Energy conservation implies that eventually the gas cools
by radiation.
As with mixing (Bayer-Kim et al.\ 2002, their section 8),
the large ratio of the number of $H\alpha$ photons to the
number of cooling hydrogen atoms is explained by this scenario.
As evident from Figure 2, there is a reasonable agreement between the
observed H$\alpha$ luminosities and the values we estimated
based on our proposed scenario.
To a certain extent, the model we have proposed reconciles the
standard cooling flow model with the recent observations of weak
X-ray line emission from gas below $\sim 10^7$ K.
If conductive cooling to cold clouds is important, these observations
might still be in accord with a moderate cooling flow model,
and cooling at such lowered rates could explain the existence
of cooler gas and young stars in the same central regions of
cooling flows, and provide a source of fuel for the central
AGNs often found there.

\acknowledgements
We benefited from discussions with Ray White at early stages of this research.
While at the University of Virginia, N. S. was supported by a Celerity
Foundation Distinguished Visiting Scholar grant.
N. S. was supported in part by grants from the
US-Israel Binational Science Foundation.
E. L. B. and C. L. S. were supported in part by
the National Aeronautics and Space
Administration through {\it Chandra} Award Numbers
GO0-1158X
and
GO1-2133X,
issued by the {\it Chandra} X-ray Center,
which is operated by the Smithsonian Astrophysical Observatory for and on
behalf of NASA under contract NAS8-39073.
Support for E. L. B. was provided by NASA through the {\it Chandra}
Fellowship
Program, grant award number PF1-20017, under NASA contract number
NAS8-39073.

\begin{deluxetable}{cccccc}
\tablewidth{0pt}
\tablecaption{Measured and Predicted $L_{H\alpha}$ for Cooling Flow Clusters \label{tab:halph}}
\tablehead{
\colhead{Cluster} & \colhead{$\dot{M}$ (T/2 to T/4)} & \colhead{kT} &
\colhead{$L_{H\alpha}$(meas.)} & \colhead{ref} & \colhead{$L_{H\alpha}$(pred.)}
\\
\colhead{} & \colhead{($M_{\odot}$ yr$^{-1}$)} &  \colhead{(keV)}
& \colhead{($\times 10^{40}$ erg s$^{-1}$)} & \colhead{} &
\colhead{($\times 10^{40}$ erg s$^{-1}$)}}

\startdata

A1835 & $800\pm200$ & $9.5\pm0.5$ & 960 & A95 & $1500\pm370$ \\
A1795 & $<80$ & $5.5\pm0.5$ & 19 & A95 & $<85$ \\
Hydra A & $120\pm60$ & $6.0\pm0.3$ & 9.4 & H89 & $140\pm70$ \\
Ser 159-03 & $<60$ & $3.8\pm0.3$ & 3.8 & H89 & $<44$ \\
2A0335+096 & $40\pm20$ & $3.2\pm0.3$ & 63 & H89 & $25\pm13$ \\
A496 & $25\pm10$ & $4.7\pm0.3$ & 1.9 & H89 & $23\pm9$ \\
MKW 3s & $<20$ & $3.7\pm0.3$ & 1.8 & H89 & $<14$ \\
A2052 & $15\pm5$ & $3.4\pm0.3$ & 2.1 & H89 & $9.8\pm3.4$ \\
A262 & $1.6\pm0.5$ & $2.1\pm0.2$ & 1.6 & H89 & $0.6\pm0.2$ \\
M87/Virgo & $0.6\pm0.2$ & $2.0\pm0.1$ & 0.6 & H89 & $0.2\pm0.08$ \\

\enddata

\tablecomments{The mass-deposition ($\dot{M}$) rates in the range 1/2 to
1/4 of the outer cluster temperature are from the {\it XMM-Newton}
high-resolution spectroscopic results of Peterson et al.\ (2003).  The
cluster temperatures [outside of the cooling flow regions, column (3)] are
also from Peterson et al.\ (2003).  The measured $H\alpha$ luminosities are
taken from either Allen (1995; A95) or Heckman et al.\ (1989; H89),
as noted in column (5).  All luminosities have been converted to be
consistent with an $H_{\circ}=70$ km s$^{-1}$ Mpc$^{-1}$, $\Omega_{M}=0.3$,
$\Omega_{\Lambda}=0.7$ cosmology.}

\end{deluxetable}

%\newpage

\begin{figure}
%\plotone{cloud.epsi}
\plotone{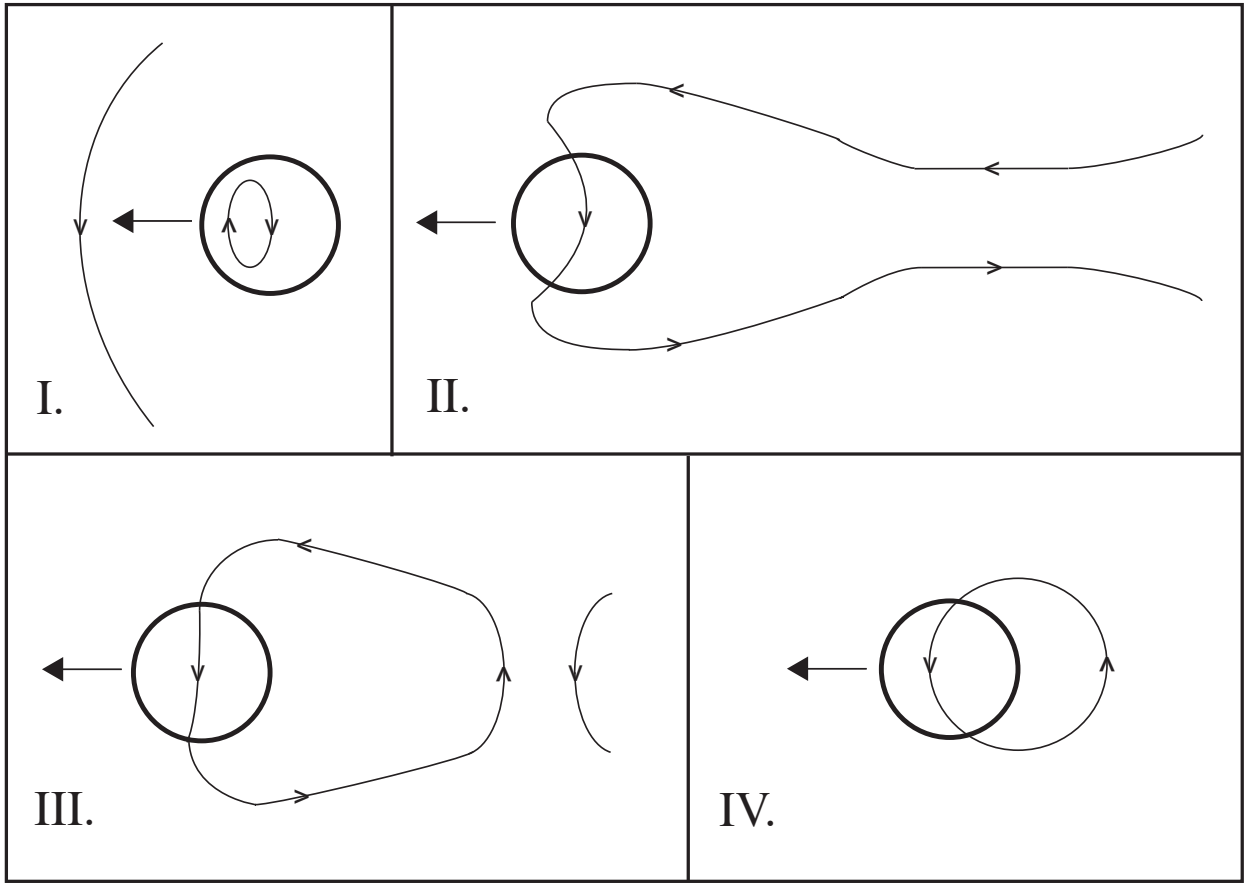}
\caption{A schematic evolution of a single intracluster magnetic field line.
The cold cloud moves to the left (panel I), such that the ambient field
line collides with the magnetic field inside the cloud.
If the two fields are not aligned (they don't need to be exactly
opposite), reconnection occurs near the front of the cloud (the
stagnation point), forming one field line threading through the cloud
and the ICM (panel II).
Because of ICM material flowing away from the region behind the
cloud (see below), the field line closes on itself behind the cloud
(panel II), leading to reconnection on the down-stream side,
and the formation of a loop connecting the cloud and ICM (panel III).
Because only a limited volume of the hot ICM is connected to the
cold cloud, heat conduction cools the ICM, and the external
pressure and magnetic stress pull the conductively-cooling
ICM toward the cloud (to the left in panel IV).}
\end{figure}

%\newpage

\begin{figure}
\plotone{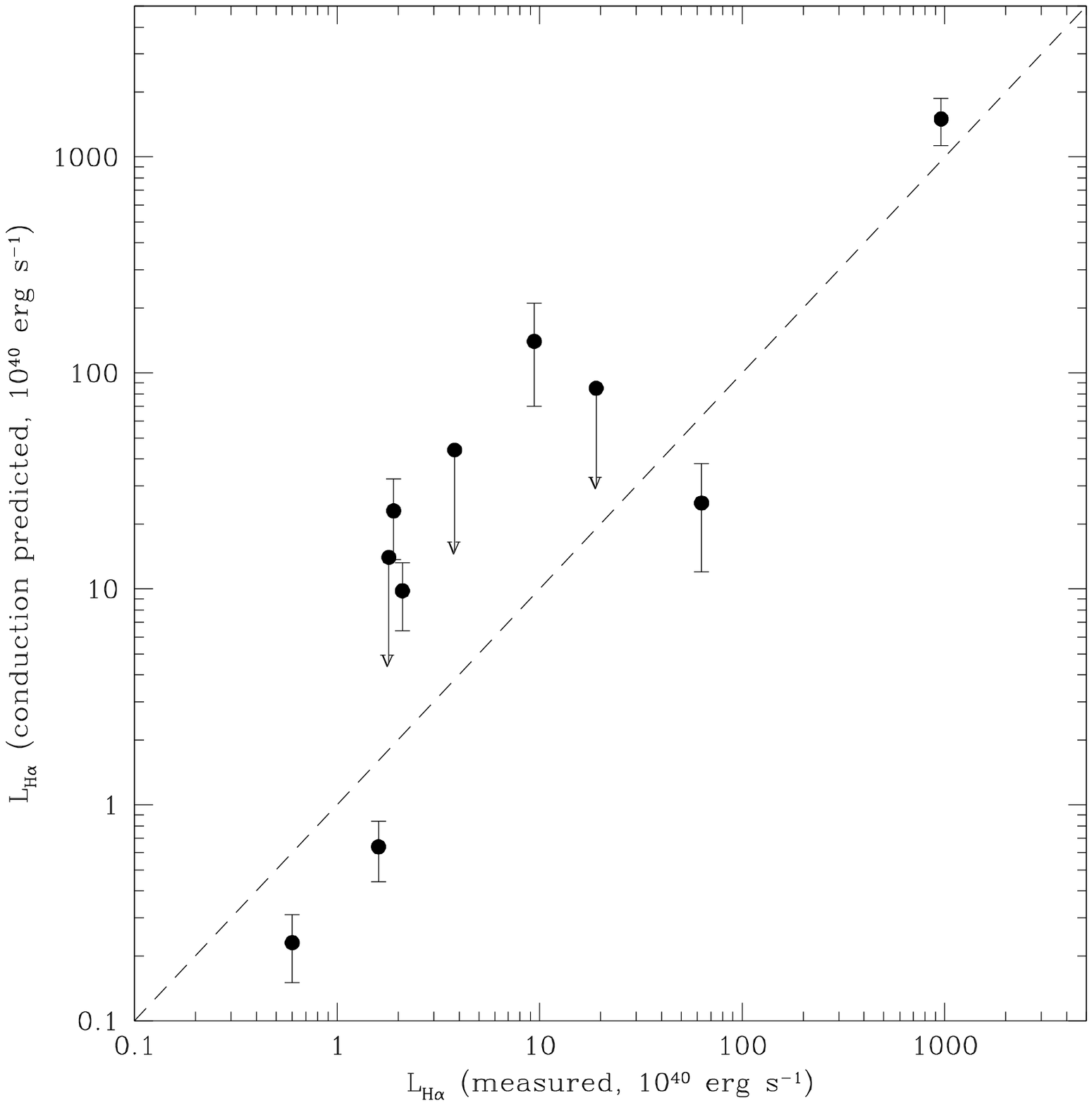}
\caption{Correlation of the measured and predicted $H\alpha$ luminosities
for the ten cooling flow clusters listed in Table 1.  The dotted line
represents a one-to-one correspondence.  Error bars are at the 1-$\sigma$
level.}
\end{figure}


\begin{references}

\reference{} Allen, S. W. 1995, MNRAS, 276, 947

\reference{} Allen, S. W., Fabian, A. C., Johnstone, R. M.,
Arnaud, K. A., \& Nulsen, P. E. J. 2001, MNRAS, 322, 589

\reference{} Balbus, S. A. 1986, ApJ, 304, 787

\reference{}  Bayer-Kim C. M., Crawford C. S., Allen S. W.,
Edge A. C., Fabian, A. C. 2002, MNRAS, 337, 938

\reference{} Begelman, M. C., \& Fabian, A. C. 1990, MNRAS, 244, 26.

\reference{} Blanton, E. L., Sarazin, C. L., McNamara, B. R., \&
Wise, M. W. 2001, ApJ, 558, L15 %  (BSMW)

\reference{} Blanton, E. L., Sarazin, C. L., \& McNamara, B. R. 2003, ApJ,
585, 227

\reference{} B\"ohringer, H., \& Fabian, A. C. 1989, MNRAS, 237, 1147

\reference{} B\"ohringer, H., Matsushita, K., Churazov, E.,
  Finoguenov, A. \& Ikebe, Y. 2004, A\&A Letters, in press
    (astro-ph/0402216)

\reference{} Borkowski, K. J.,  Balbus, S. A., \& Fristrom, C. C. 1990,
    ApJ, 355, 501

\reference{} Bregman, J. N., \& David, L. P. 1988, ApJ, 326, 639

% ?? mew referemce

\reference{} Cho, J., Lazarian, A., Honein, A., Knaepen, B., Kassinos, S.,
Moin, P. 2003, ApJ, 589,  L77

\reference{} Edge A. C. 2001, MNRAS, 328, 762

\reference{} Edge A. C., \& Frayer, D. T. 2003, ApJ, 594, L13

% ?? this is almost certainly not the reference you intended
% \reference{} Edge, A. C., Wilman, R. J., Johnstone, R. M.,
%    Crawford, C. S., Fabian, A. C., \& Allen, S. W. 2002, MNRAS, 337, 49

\reference{} Eilek, J. A., \& Owen, F. N. 2002, ApJ, 567, 202

\reference{} Fabian, A. C. 2002, in Galaxy Evolution: Theory and
Observations, ed.\ V. Avila-Reese, C. Firmani, C.
Frenk, \& C. Allen,  RevMexAA SC, in press (astro-ph/0210150)

\reference{} Fabian, A. C., Allen, S. W., Crawford, C. S.,
    Johnstone, R. M., Morris, R. G., Sanders, J. S., \& Schmidt, R. W.
     2002a, MNRAS, 332, L50

\reference{} Fabian, A. C.,  Mushotzky, R. F.,  Nulsen, P. E. J.,
\&  Peterson, J. R. 2001a, MNRAS, 321, L20

\reference{} Fabian, A. C., Sanders, J. S., Ettori, S., Taylor, G. B.,
Allen, S. W., Crawford, C. S., Iwasawa, K., \& Johnstone, R. M. 2001b,
MNRAS, 321, L33

\reference{} Fabian, A. C., Voigt, L. M., \& Morris, R. G. 2002b,
MNRAS, 335, L71

\reference{} Ferland, G. J., Fabian, A. C., \& Johnstone, R. M.
    2002, MNRAS, 333, 876

\reference{} Friaca, A. C. S., Goncalves, D. R., Jafelice, L. C.,
     Jatenco-Pereira, V., \& Opher, R. 1997, A\&A, 324, 449

\reference{} Gaetz, T. J., Edgar, R. J., \& Chevalier, R. A.
            1988, ApJ, 329, 927

\reference{} Godon, P., Soker, N., \& White, R. E., III 1998, AJ, 116, 37

\reference{} Gregori, G., Miniati, F., Ryu, D., \& Jones, T. W. 2000
  ApJ, 543, 775

\reference{} Heckman, T. M., Baum, S. A., van Breugel, W. J. M., \& McCarthy, P. J.
1989, ApJ, 338, 48

\reference{} Jafelice, L. C., \& Friaca, A. C. S. 1996, MNRAS, 280, 438

\reference{} Johnstone, R. M., Allen, S. W., Fabian, A. C., \&
  Sanders, J. S. 2002, MNRAS, 336, 299

\reference{} Kaastra, J. S., Ferrigno, C., Tamura, T., Paerels, F. B. S.,
     Peterson, J. R., \& Mittaz, J. P. D 2001, A\&A, 365, L99

\reference{} Lazarian, A., \& Cho, J. 2003, in JENAM2002,
The Unsolved Universe: Challenges for the Future (astro-ph/0302104)

\reference{} Loeb, A. 2002, NewA, 7, 279

\reference{} Loewenstein, M., \& Fabian, A. C. 1990, MNRAS, 242, 120

\reference{} Markevitch, M., Vikhlinin, A., Forman, W. R. 2002, in
``Matter and Energy in Clusters of Galaxies",
to appear in ASP Conference Series (astro-ph/0208208)

\reference{} McKee, C. F., \& Begelman, M. C. 1990, ApJ, 358, 392

\reference{} McNamara, B. R. et al.\ 2000, ApJ, 534, L135

\reference{} McNamara, B. R. 2002, The High-Energy Universe at
Sharp Focus: Chandra Science, Symposium at the ASP meeting, in press
(astro-ph/0202199)

\reference{} Miniati, F.,  Jones, T. W., \& Ryu, D. 1999,
 ApJ, 517, 242

\reference{} Molendi, S., \& Pizzolato, F. 2001, ApJ, 560, 194

\reference{} Morris, R. G., \& Fabian, A. C. 2003, MNRAS, 338, 824

\reference{} Narayan, R., \& Medvedev, M. V. 2001, ApJ, 562, L129

\reference{} Norman, C.,  \& Meiksin, A. 1996, ApJ, 468, 97

\reference{} Oegerle, W. R., Cowie, L., Davidsen, A., Hu, E.,
   Hutchings, J., Murphy, E., Sembach, K., \& Woodgate, B.
   2001, ApJ, 560, 187

\reference{} Oieroset, M., Phan, T. D., Fujimoto, M., Lin, R. P.,
     \& Lepping, R. P. 2001, Nature, 412, 414

\reference{} Peterson, J. R.,  et al.\ 2001, A\&A, 365, L104

\reference{} Peterson, J. R., Ferrigno, C., Kaastra, J. S., Paerels,
F. B. S., Kahn, S. M., Jernigan, J. G., Bleeker, J. A. M., \& Tamura, T.
2002, in New Visions of the X-ray Universe in the
XMM-Newton and Chandra Era, in press (astro-ph/0202108)

\reference{} Peterson, J. R., Kahn, S. M., Paerels, F. B. S., Kaastra, J.
S., Tamura, T., Bleeker, J. A. M., Ferrigno, C., \& Jernigan, J. G. 2003,
ApJ, 590, 207

\reference{} Petschek, J. E., \& Thorne, R. M. 1967, ApJ, 147, 1157

\reference{} Pistinner, S. L., \& Eichler, D. 1998, MNRAS, 301, 49

\reference{} Pistinner, S. L., Levinson, A., \& Eichler, D. 1996,
        ApJ, 467, 162

\reference{} Pistinner, S., \& Shaviv, G. 1996, ApJ, 459, 147

\reference{} Ruszkowski, M., \& Begelman, M. C. 2002, ApJ, 581, 223

\reference{} Sabra, B. M., Shields, J. C., \& Filippenko, A. V.
  2000, ApJ, 545, 157

\reference{} Salome, P., \& Combes, F. 2003, A\&A in press (astro-ph/0309304)

\reference{} Soker, N. 1997, ApJ., 488, 572

\reference{} Soker, N., \& Dgani, R. 1997, ApJ, 484, 277

\reference{} Soker, N., Sarazin, C. L. 1990, ApJ, 348, 73

\reference{} Soker, N., White, R. E. III, David, L. P., \& McNamara, B. R.
    2001, ApJ, 549, 832

\reference{} Taylor, G. B., Fabian, A. C., \&  Allen, S. W. 2002,
MNRAS, 334, 769

{\bf
\reference{} Vikhlinin, A., Markevitch, M., \& Murray, S. S. 2001, ApJ,
549, L47
}

\reference{} Voigt, L. M., Schmidt, R. W., Fabian, A. C., Allen, S. W.,
   \& Johnstone, R. M. 2002, MNRAS, 335, L7

\reference{} Young, A. J., Wilson, A. S., \& Mundell, C. G.
2002, ApJ, 579, 560

\end{references}
\end{document}